\documentclass[12pt,a4paper]{article}

\begin{document} 

\hfill {}
\vskip 0.7 true cm
\begin{center}
{\Large \bf Several remarks on ``Comments'' by A. Moroz}
\end{center}
\vskip 1.5 true cm
%% def of soft t; t'=\symbol{187}U
%%\def\Ht{\mbox{t\kern-
%%.35ex\protect\raisebox{1.3ex}[0ex][0ex]{$,$}\kern-.1ex}}
\begin{center}
%{\large P. \v S{\Ht}ov\'{\i}\v cek}
\textsc{\large P. \v S{\v t}ov\'{\i}\v cek}
%% P. Stovicek
\end{center}
\begin{center}
{\it Department of Mathematics, Faculty of Nuclear Science, CTU,\\
       Trojanova 13, 120 00 Prague, Czech Republic
}
\end{center}
\vskip 2 true cm

\begin{center}
{\large Abstract}
\end{center}
\bigskip
\noindent
\hspace{.5in}\begin{minipage}{5in}
We make a couple of remarks on ``Comments'' due to A. Moroz  
which were addressed to our recent letter \cite{1}.
\end{minipage}

\vskip 1.5 true cm

\begin{minipage}{4.5in}
{PACS. 03.65.Nk -- Nonrelativistic scattering theory }
\end{minipage}
\vskip 1.5 true cm

\newpage

\noindent 
In this note we wish to make a couple of remarks on ``Comments on 
'Differential cross section for Aharonov--Bohm effect with non-standard 
boundary conditions' `` (referred to as Comments in what follows) 
due to A. Moroz and addressed to our recent letter \cite{1}. 
In fact, these remarks have been revised as a consequence of 
the revision of Comments. 
In the very beginning it should be emphasized that in our letter \cite{1} 
we didn't pretend and attempt to do anything more than claimed in the 
introductory part, and this was to complete the results of the 
preceding paper \cite{2} 
by doing some elementary numerical analysis, plotting a couple of graphs, 
and providing them with some basic discussion. However all the involved 
formulae were derived in \cite{2} (of course, the rotational symmetry was 
mentioned there, too). 
The basic result of \cite{2} consists in 
finding a general form of boundary conditions, depending on four parameters, 
imposed on a wave function in the case of idealized Aharonov--Bohm effect 
in the plane, and furthermore in derivation of a formula for the 
differential cross section with the corresponding Hamilton operator. 
The value of the magnetic flux is then another parameter, 
the fifth one, occurring in the problem. 

It should be noted that approximately at the same time another paper 
appeared, \cite{3}, treating the same problem as we did in \cite{2}, 
using practically the same 
mathematical tools, and naturally arriving at the same results. 
So Comments by Moroz might have been better addressed to the papers 
\cite{2} and \cite{3} rather than to \cite{1}. 
 
The mathematical machinery applied in \cite{2} and \cite{3} is based on the 
theory of self-adjoint extensions of symmetric operators which is nowadays 
quite a common tool suited for this type of problems. For example, just 
for the purpose of illustration, let us mention that in another relatively 
recent paper, \cite{4},  
in its nature a similar problem, concerning this time a magnetic monopole 
in three-dimensional space, has been solved by the same method. 
 
As I understood from Comments, Moroz had applied in his analysis 
a completely different technique based on a limit procedure when sending 
the radius of the flux tube to zero. There is no doubt that this approach 
is highly interesting, too. For example one may hope to give 
this way the involved 
parameters a more concrete interpretation. Unfortunately I am not able to  
compare the results directly 
since I didn't deduce from Comments what 
was the precise definition of Hamiltonian, particularly what kind of 
boundary conditions were imposed on wave functions. Nevertheless some 
aspects seem to be clear. First of all, Comments are apparently concerned 
with the case when the s and p-wave are decoupled. Then the corresponding 
family of Hamiltonians depends only on two parameters, called 
$\Delta_{-n}$ and $\Delta_{-n-1}$ in formula (3) of Comments, 
while the complete solution admits coupling of the s and p-wave via boundary 
conditions, and consequently depends, as already mentioned, 
on four parameters. 

As far as the rotational symmetry is concerned, this notion is interpreted 
somewhat differently in \cite{2,3} on one side and in Comments on the 
other side. 
This was not our aim to study symmetries of the 
differential cross section itself, for example with respect to the 
reflection $\varphi\to-\varphi$. A basically more essential observation 
has been made in \cite{2,3} (and graphically illustrated in \cite{1}). 
The point is 
that the angular momentum need not be conserved, or, in other words, 
the angular momentum need not commute with the Hamiltonian. This effect 
happens provided the s and p-wave are coupled, and this is what we called 
the violation of the rotational symmetry. For the differential cross 
section this means that it depends non-trivially on both the scattering and 
incident angle, $\varphi$ and $\varphi_0$, and not merely on their 
difference $\varphi-\varphi_0$. Note that only the scattering angle 
$\varphi$ occurs in formula (3) of Comments, with $\varphi_0$ being 
set to 0. 

Furthermore, let us make a short remark on the number of bounded states. 
The situation is slightly more complicated than mentioned in Comments. The 
possible number of bounded states is 0, 1 or 2. This means that even with 
non-standard boundary conditions it may happen that there are 
no bounded states. The 
dependence of the number of bounded states on the choice of boundary 
conditions has been analyzed quite explicitly both 
in \cite{2} and \cite{3}. 

As for the longer history, the fact that the two critical sectors of 
angular momentum admit more general boundary conditions than the regular 
one has been known for a long time, see for example the now classical 
paper \cite{5},  
though this possibility was not exploited systematically until recently. 
We have to admit that we missed the work \cite{6}.  
No doubt it should be included among the references of \cite{1}. 
On the other hand I'd like 
to point out that the case with the s and p-wave decoupled was studied 
in several works prior to the paper \cite{7} referred to in Comments   
(see \cite{8,9}). 

%%%%%%%%%%%%%%%%%%%%%%%%%%%%%%
%\vskip 12pt
%\noindent\textbf{Note added in proof.} 
%These remarks were written as a response 
%to the original version of Comments 
%by A. Moroz and this is why some slight 
%discrepancies may be found with 
%respect to the updated text of Comments. 

%%%%%%%%%%%%%%%%%%%%%%%%%%%%%%
%\newpage

\end{document}